\documentclass[
 reprint,
superscriptaddress,
 amsmath,amssymb,
 aps,
pra,
]{revtex4-1}
\usepackage{color}
\usepackage{graphicx}
\usepackage{caption}
\usepackage{subcaption}
\usepackage{dcolumn}
\usepackage{bm}
\usepackage{braket}
\usepackage{float}
\usepackage{multirow}
\usepackage{mathtools}
\usepackage{qcircuit}
\begin{document}


\title{Stabilizer Slicing: Coherent Error Cancellations in LDPC Codes}

\author{Dripto M. Debroy}
\email{dripto@phy.duke.edu}
\affiliation{Department of Physics, Duke University, Durham, NC 27708, USA}
\author{Muyuan Li}
\affiliation{School of Computational Science and Engineering,
 Georgia Institute of Technology, Atlanta, Georgia 30332, USA}
 \author{Michael Newman}
 \affiliation{Department of Physics, Duke University, Durham, NC 27708, USA}
\affiliation{Departments of Electrical and Computer Engineering and Chemistry, Duke University, Durham, NC 27708, USA}
\author{Kenneth R. Brown}
\email{ken.brown@duke.edu}
\affiliation{Department of Physics, Duke University, Durham, NC 27708, USA}
\affiliation{School of Computational Science and Engineering,
 Georgia Institute of Technology, Atlanta, Georgia 30332, USA}
\affiliation{Departments of Electrical and Computer Engineering and Chemistry, Duke University, Durham, NC 27708, USA}

\begin{abstract}
Coherent errors are a dominant noise process in many quantum computing architectures. Unlike stochastic errors, these errors can combine constructively and grow into highly detrimental overrotations. To combat this, we introduce a simple technique for suppressing systematic coherent errors in low-density parity-check (LDPC) stabilizer codes, which we call \emph{stabilizer slicing}.  The essential idea is to slice low-weight stabilizers into two equally-weighted Pauli operators and then apply them by rotating in opposite directions, causing their overrotations to interfere destructively on the logical subspace.

With access to native gates generated by $3$-body Hamiltonians, we can completely eliminate purely coherent overrotation errors, and for overrotation noise of $0.99$ unitarity we achieve a $135$-fold improvement in the logical error rate of Surface-$17$.  For more conventional $2$-body ion trap gates, we observe an $89$-fold improvement for Bacon-Shor-$13$ with purely coherent errors which should be testable in near-term fault-tolerance experiments.  This second scheme takes advantage of the prepared gauge degrees of freedom, and to our knowledge is the first example in which the state of the gauge directly affects the robustness of a code's memory. This work demonstrates that coherent noise is preferable to stochastic noise within certain code and gate implementations when the coherence is utilized effectively.

\end{abstract}

\pacs{Valid PACS appear here}
\maketitle
As we grow closer to experimentally implementing small quantum error-correcting codes~\cite{BermudezITQCwithSteane2017, trout2018simulating,  OBrienSurface17DensityMatrix2017}, it becomes increasingly important to study physically motivated error models. While extensive work has been done on simulating the behavior of codes under Pauli noise~\cite{KnillKnillEC2005, RaussendorfClusterState12007, muyuan2018direct,li20182}, other more realistic models have drawn less attention.

Coherent errors are small unitary operations that occur consistently after every gate, and often require different tools to correct~\cite{GutierrezIncohCohNoiseEC2016,darmawan2017tensor,suzuki2017efficient,bravyi2017correcting,chamberland2017hard}. These errors can arise from a variety of sources, and are generally more damaging due to their capacity to grow rapidly when combined~\cite{huang2018performance,kueng2016comparing,greenbaum2017modeling,beale2018coherence}. In ion trap quantum computing, coherent errors are a dominant noise source, stemming from miscalibrations in experimental equipment, such as laser intensity. This systematic error leads to a relative angle of overrotation, and is a larger source of infidelity than decoherence in most cases. 

The intensity of the laser drifts slowly relative to gate times, and so the fraction of overrotation remains approximately constant over a given error correction cycle. This is a double-edged sword: while these errors rapidly accumulate as they are repeatedly applied, they also have a predictable form. Existing proposals take advantage of this uniformity to suppress coherent errors~\cite{brown2004arbitrarily}. However these are commonly restricted to improving single-qubit gates due to excessively long gate times when applied to multi-qubit gates~\cite{jones2003robust,tomita2010multi}.  

Although we draw our motivation from ion trap quantum computing, the systematic inaccuracy of multi-qubit gates is a major bottleneck for many different architectures. Thus, mitigating coherent errors is an important problem which would benefit any architecture.

Our method takes advantage of experimental degrees of freedom to suppress systematic coherent errors in one of the most important fault-tolerance circuits: syndrome extraction.  Importantly, we can reduce the logical error rates without improving the constituent gates or requiring any additional overhead.

To do so, we introduce a technique which we call \emph{stabilizer slicing}.  The essential idea is to split a stabilizer into two equally weighted Pauli rotations, and apply them in opposite directions.  In this way, systematic overrotations destructively interfere, leaving only the intended gates.

\section*{Architectural Requirements}

To perform stabilizer slicing, we require a quantum computing architecture with two particular experimental degrees of freedom.
\begin{itemize}
    \item[$(i)$] Our architecture gives us the directional freedom to apply any gate in the clockwise or counterclockwise direction.
    \item[$(ii)$] For a code with $2n$-body stabilizers, our architecture can generate native multi-qubit gates by evolving an $(n+1)$-body Hamiltonian.
\end{itemize}

\begin{figure}[t!]
    \centering
    \resizebox{.65\linewidth}{!}{
    \mbox{
    \Qcircuit @C=.9em @R=.7em {
     & \gate{H} & \qw & \push{\rule{.3em}{0em}=\rule{.3em}{0em}} & & \gate{RX(-\pi)} & \gate{RY(\frac{\pi}{2})} & \qw\\
    & & & \push{\rule{.3em}{0em}=\rule{.3em}{0em}} & & \gate{RY(-\frac{\pi}{2})} & \gate{RX(\pi)} & \qw
    }
    }}
    \vspace{1.5em}
    
    \resizebox{\linewidth}{!}{
    \Qcircuit @C=.9em @R=.7em {
     & \ctrl{1} & \qw & \raisebox{-2.5em}{=} & & \gate{RY(v\frac{\pi}{2})} & \multigate{1}{XX(s\frac{\pi}{4})} & \gate{RX(-s\frac{\pi}{2})} & \gate{RY(-v\frac{\pi}{2})} & \qw\\
   & \targ & \qw & & & \qw & \ghost{XX(s\frac{\pi}{4})} & \gate{RX(-vs\frac{\pi}{2})} & \qw & \qw
    }}
    \caption{Ion trap gate compilations of CNOT and $H$ in terms of one- and two-qubit Pauli rotations~\cite{MaslovCircuitCompIT2017}. The choices of $s,v \in \{\pm 1\}$ represent rotational degrees of freedom.}
    \label{gates}
\end{figure}
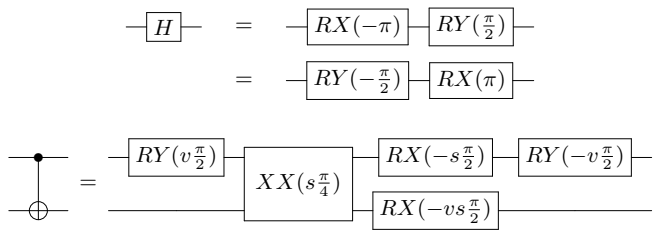

Even when we restricted to standard $2$-body interactions (i.e. $n = 1$), our technique has near-term applications to Shor's code, and to the boundary of surface codes. The directional freedom in $(i)$ can be seen for ion trap multi-qubit M\o{}lmer-S\o{}renson gates applied to ions of varying interaction parameters, see Figure~\ref{gates} \cite{MaslovCircuitCompIT2017}.  In this case, the freedom of direction can be realized experimentally by adjusting the relative phase of the Raman beams driving the entangling gate.
When restricting to low-weight stabilizers in LDPC codes, conditions similar to $(ii)$ are already being proposed for directly implementing multi-qubit measurements in super-conducting systems \cite{divincenzo2013multi}; coupled with subsystem surface code constructions, the reduced circuit volume of syndrome extraction can yield significantly higher thresholds \cite{bravyi2012subsystem}. 

Given an architecture that satisfies $(i)$ and $(ii)$, we can implement syndrome extraction circuits which are robust to coherent errors.  The insight is to use $(i)$ and $(ii)$ to direct corresponding overrotations against each other, thus cancelling them.
\section*{Stabilizer Slicing}

To illustrate this simple technique, we first consider a leading example. Imagine you are challenged to create the highest fidelity identity channel possible using two applications of a gate $G$ which satisfies $G^2 = I$. Since $G$ is an involution, we can express it as:
\begin{equation}
    \exp(-i \theta G) = \cos(\theta)I - i \sin(\theta) G
    \label{gaterot}
\end{equation}
where G is applied when $\theta = \pm\pi/2$. In the noiseless case, applying two positive rotations is the same as applying one positive and one negative, up to global phase. However if there exists some coherent overrotation due to miscalibrations in your experimental setup, these cases diverge. In the case where both rotations are in the same direction, errors of the form $\pi/2 \rightarrow (1+\epsilon)\pi/2$ add constructively, while in the case where the gates are applied in opposite directions, the errors destructively interfere.

\begin{figure}[t!]
\includegraphics[width = \linewidth]{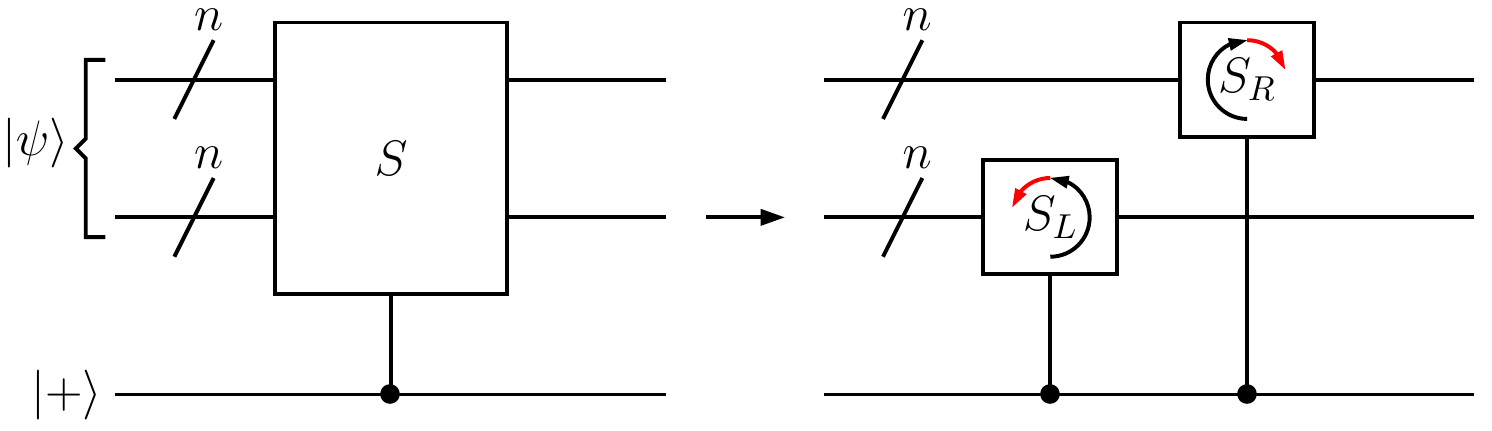}
\caption{Stabilizer slicing in bare-ancilla syndrome extraction.  In this case, $S = S_LS_R$ are disjoint and $\ket{\psi}$ is a clean codestate so that $S\ket{\psi} = \ket{\psi}$.  The controlled-$S_L$ and controlled-$S_R$ are performed using opposite rotations.  The relative overrotations (red) cancel on the logical subspace.}
\label{stabilizer_slicing}
\end{figure}

While identity circuits don't come up often, the intuition is similar for our cancellations from stabilizer slicing. When acting on a clean codestate, a stabilizer $S$ effectively acts as an identity operator. Suppose the stabilizer is split into two evenly weighted rotations $S_L$ and $S_R$ satisfying $S_L S_R = S$. Then, if $|\psi\rangle$ is a clean codestate,
\begin{equation}
    \begin{split}
        S|\psi\rangle &= |\psi\rangle\\
        S_R |\psi\rangle &= S_L |\psi\rangle\\
        \therefore \exp(i\theta S_R)|\psi\rangle &= \exp(i\theta S_L)|\psi\rangle.
    \end{split}
    \label{slsr}
\end{equation}
Following a similar intuition as the previous case, we can apply our stabilizer through two controlled $\pi/2$-rotations. If our errors are overrotations of the form 
\begin{equation}
    \begin{split}
        U_{E}^{L} := \exp(i\epsilon\theta_L CS_L)\\
        U_{E}^{R} := \exp(i\epsilon\theta_R CS_R),
    \end{split}
    \label{errdef}
\end{equation}
 then by Equation \ref{slsr} we can replace $S_R$ with $S_L$ and have $\theta_L$ and $\theta_R$ point in opposite directions in order to completely cancel our errors. 
 \begin{equation}
     \begin{split}
           U_{E}^{L}U_{E}^{R}|\psi\rangle |+\rangle = &\exp(i\epsilon\theta_L CS_L)\exp(i\epsilon\theta_R CS_R)|\psi\rangle |+\rangle\\
         = &\exp(i\epsilon\theta_L CS_L)\exp(i\epsilon (-\theta_L) CS_L) |\psi\rangle |+\rangle\\
         = &|\psi\rangle |+\rangle.
     \end{split}
     \label{errcancel}
 \end{equation}
 This technique of applying a stabilizer in two halves which have perfectly cancelling errors is what we call \emph{stabilizer slicing}, and the corresponding circuit is shown in Figure \ref{stabilizer_slicing}.
Of course, we perform syndrome extraction to detect errors, and so sometimes $\ket{\psi}$ will \emph{not} be a clean codestate. If some stochastic error $E$ occurs on our data, it may put the data into a state for which $S|\psi\rangle = -|\psi\rangle$. In this case, a stabilizer sliced circuit would actually grow coherent errors when measuring violated stabilizers. In this way, we can see that coherent error suppression via stabilizer slicing interfaces non-trivially with other sources of noise. However, in the low-error regime, the majority of stabilizers will commute with $E$ and so stabilizer slicing will have an overwhelmingly positive effect on coherent errors in total.

It is worth noting why property $(ii)$ is necessary. Although we can express any such controlled-$S$ as a product of noiseless two-qubit gates, overrotations on the individual gates will not perfectly cancel.  For a weight-$n$ stabilizer $S$ expanded as a product of $m \geq 3$ multi-qubit rotations, interference only occurs between weight $k$ and weight $n-k$ components. Near $\theta_i = \pm \pi/2$, this yields only a negligible suppression of the two-qubit coherent errors.  Thus, splitting the stabilizer into precisely two native components is essential for perfect cancellation.

\section*{Simulation Details and Results}
\begin{figure}[t!]
    \centering
    \includegraphics[width = .9\linewidth]{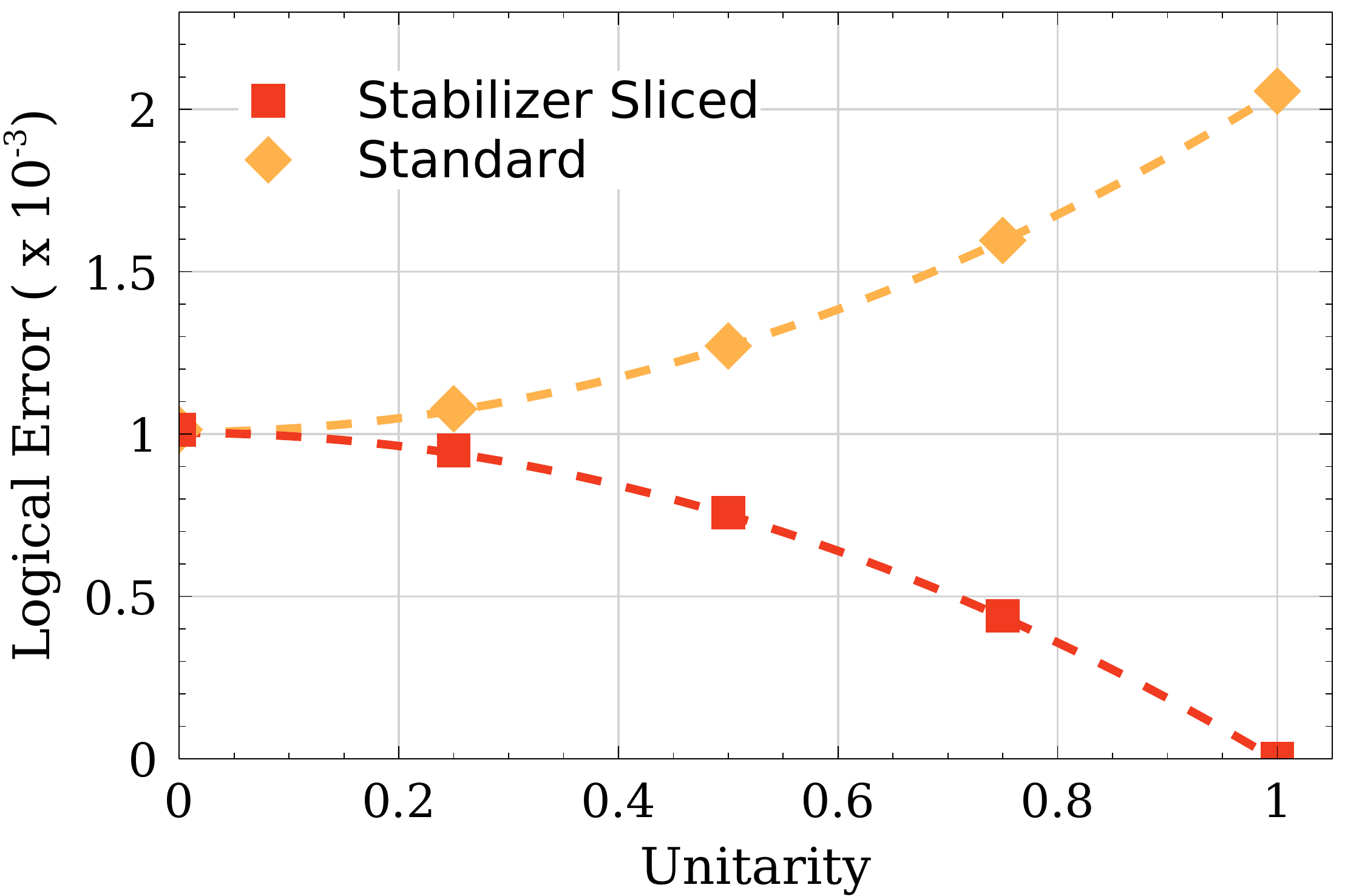}
    \caption{Logical error rates and quadratic fits for Surface-17 assuming access to native $3$-body gates, both with and without stabilizer slicing. As expected, in the fully coherent case we completely eliminate the noise present in our model. In this plot, the $2$- and $3$-qubit gate infidelities are $1.0 \times 10^{-3}$.}
    \label{surface_3_body}
\end{figure}
There are two central simulations presented in this paper. The first is a simulation of Surface-17~\cite{SCorig, Tilted13SC, versluis2017scalable, TomitaLowDSC2014, fowler2012surface}, where we allow ourselves native $3$-body operations and which features a simplified error model. This simulation is intended to show the performance of stabilizer slicing without any other confounding factors. Our second simulation is for a $13$ qubit Bacon-Shor code~\cite{BaconBaconShor2006,ShorBaconShor1995,aliferis2007subsystem} written in the native ion trap gate set, which shows how our technique would perform on a near-term machine. This second simulation has a physically motivated overrotation error model based on a miscalibration in laser intensity. An error model that includes dephasing noise is included in the Supplemental Materials, along with one with SK1 implemented~\cite{brown2004arbitrarily}.

To better understand stabilizer slicing as errors transition from coherent to stochastic, we model our overrotation errors as a mixed channel with two parameters, as this is a better model of physical errors~\cite{wallman2015estimating,sheldon2016characterizing,laflamme2016mixed}. The first parameter is the unitarity $\kappa$. The second parameter is the overrotation angle $\epsilon$, which controls the strength of the error. Consequently, the error following some perfect gate $G$ has the form,
\begin{equation}
    \varepsilon_G(\rho) = \kappa\cdot\varepsilon_G^c(\rho) + (1 - \kappa)\cdot\varepsilon_G^s(\rho)
    \label{errchan}
\end{equation}
where $\varepsilon_G^c$ and $\varepsilon_G^s$ are coherent and stochastic overrotation channels with equal fidelity given by,
\begin{equation}
    \begin{split}
        \varepsilon_G^c(\rho) &= \exp (-i\epsilon G)\rho\exp (i\epsilon G)\\
        \varepsilon_G^s(\rho) &= \cos^2(\epsilon)I\rho I + \sin^2(\epsilon)G\rho G.
    \end{split}
    \label{channels}
\end{equation}
We reiterate that, ordinarily, we would expect the latter channel, corresponding to $\kappa = 0$, to produce lower logical error rates due to the dropout of off-diagonal terms. To best display the improvements of stabilizer slicing, we do not include measurement error.
\begin{figure}[t!]
    \centering
    \includegraphics[width = .9\linewidth]{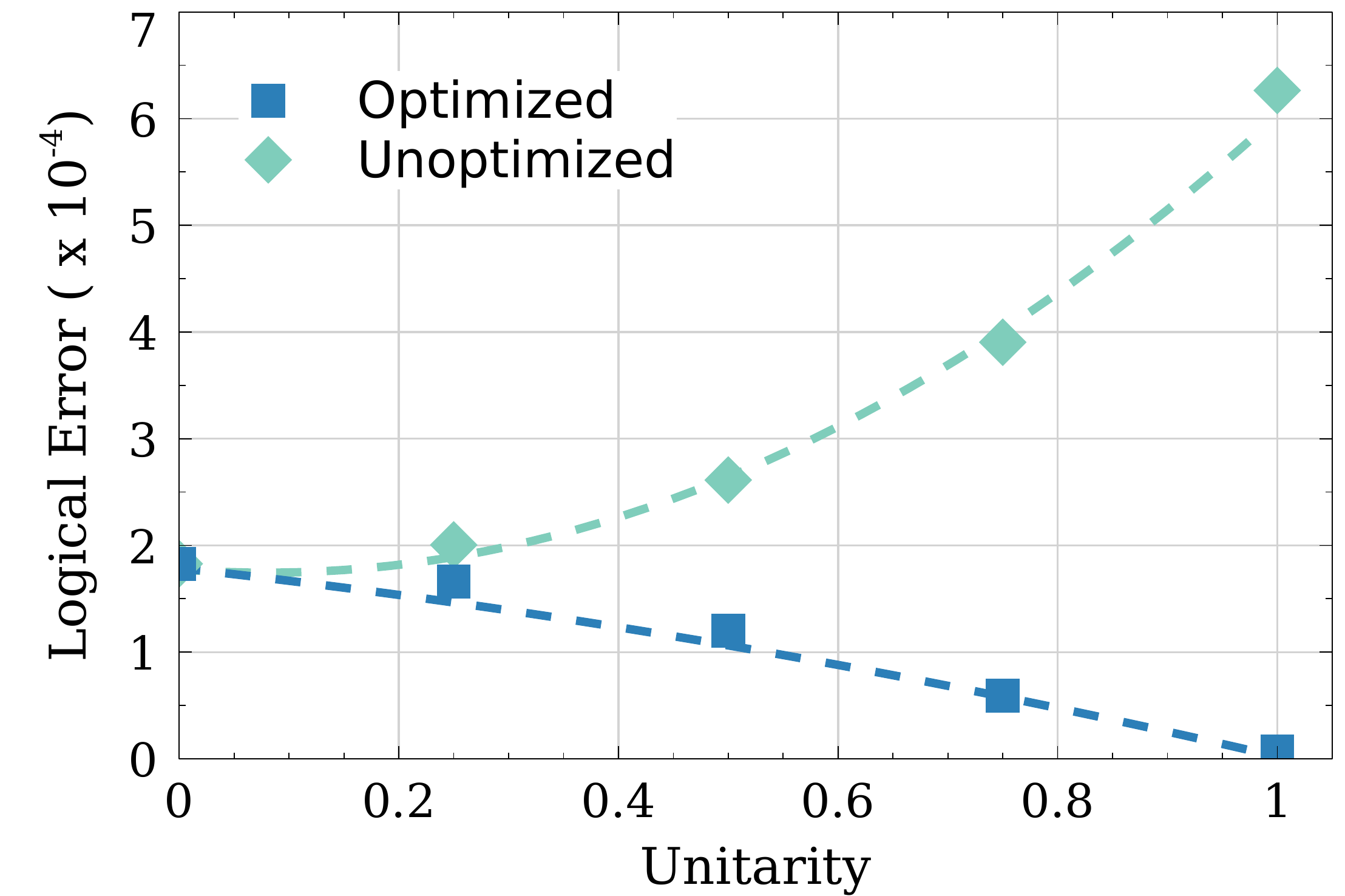}
    \caption{One-sided logical error rates for Bacon- Shor-$13$, both with and without stabilizer slicing. The optimized line does not intersect the $x$-axis in the purely coherent case since we have included single qubit overrotations. However these can be dealt with using SK1, as we describe in the Supplemental Materials. In this plot, the two-qubit gate infidelity is $\sin^2(\epsilon_2) = 5.0 \times 10^{-4}$ and $\epsilon_2 = (1+\epsilon_1)^2-1$.}
    \label{shor}
\end{figure}
Since coherent error is difficult to simulate, we must use more computationally intensive techniques. Using the full density simulator \emph{quantumsim}, we write our error correcting circuits as a full quantum channel~\cite{OBrienSurface17DensityMatrix2017}. By postselecting on all possible syndrome outcomes and then combining the resulting sub-normalized density matrices, we are able to calculate exact logical error rates. While this technique is perfectly accurate, it is exponential in the number of postselections, and consequently we are restricted to simulations with perfect preparation of logical states.

The first simulation we show is for a stabilizer sliced Surface-$17$ code. In this simulation we assume access to $3$-body $CXX$ and $CZZ$ gates, and have an error model where $\epsilon_2 = \epsilon_3$ for simplicity. It should be noted that since the $2$-body and $3$-body gates are never applied within the same stabilizer, this condition is not important for the effectiveness of our scheme. As can be seen in Figure~\ref{surface_3_body}, interaction between the stochastic and coherent errors leads to a non-linear interpolation as we move from stochastic to coherent.

\begin{figure}[ht!]
    \mbox{
    \Qcircuit @C=1.8em @R=.7em {
        & \qw & \targ & \qw & \qw & \qw & \qw & \qw & \qw & \qw\\
        & \qw & \qw & \qw & \targ & \qw & \qw & \qw & \qw & \qw\\
        & \qw & \qw & \targ & \qw & \qw & \qw & \qw & \qw & \qw\\
        & \qw & \qw & \qw & \qw & \targ & \qw & \qw & \qw & \qw\\
        & \qw & \qw & \qw & \qw & \qw & \qw & \targ & \qw & \qw\\
        & \qw & \qw & \qw & \qw & \qw & \targ & \qw & \qw & \qw\\
        \lstick{\ket{0}} & \gate{H} & \ctrl{-6} & \ctrl{-4} & \ctrl{-5} & \ctrl{-3} & \ctrl{-1} & \ctrl{-2} & \gate{H} &\meter
    }
    }
    
    \vspace{2em}
    
    \mbox{
    \Qcircuit @C=.7em @R=.5em {
        & \multigate{6}{\rotatebox[origin=c]{270}{\scriptsize $XX(+)$}} & \gate{ X(+)} &  \qw & \qw & \qw & \qw & \qw & \qw\\
        & \ghost{\rotatebox[origin=c]{270}{\scriptsize $XX(+)$}} & \qw &  \multigate{5}{\rotatebox[origin=c]{270}{\scriptsize $XX(+)$}} & \gate{\scriptsize X(+)} & \qw & \qw & \qw & \qw\\
        & \ghost{\rotatebox[origin=c]{270}{\scriptsize $XX(+)$}} & \multigate{4}{\rotatebox[origin=c]{270}{\scriptsize $XX(-)$}} &  \ghost{\rotatebox[origin=c]{270}{\scriptsize $XX(+)$}} & \gate{\scriptsize X(-)} & \qw & \qw & \qw & \qw\\
        & \ghost{\rotatebox[origin=c]{270}{\scriptsize $XX(+)$}} & \ghost{\rotatebox[origin=c]{270}{\scriptsize $XX(-)$}} & \ghost{\rotatebox[origin=c]{270}{\scriptsize $XX(+)$}} & \multigate{3}{\rotatebox[origin=c]{270}{\scriptsize $XX(-)$}} & \gate{\scriptsize X(-)} & \qw & \qw & \qw\\
        & \ghost{\rotatebox[origin=c]{270}{\scriptsize $XX(+)$}} & \ghost{\rotatebox[origin=c]{270}{\scriptsize $XX(-)$}}& \ghost{\rotatebox[origin=c]{270}{\scriptsize $XX(+)$}} & \ghost{\rotatebox[origin=c]{270}{\scriptsize $XX(-)$}} & \qw & \multigate{2}{\rotatebox[origin=c]{270}{\scriptsize $XX(-)$}} & \gate{\scriptsize X(-)} & \qw\\
        & \ghost{\rotatebox[origin=c]{270}{\scriptsize $XX(+)$}} & \ghost{\rotatebox[origin=c]{270}{\scriptsize $XX(-)$}}& \ghost{\rotatebox[origin=c]{270}{\scriptsize $XX(+)$}} & \ghost{\rotatebox[origin=c]{270}{\scriptsize $XX(-)$}} & \multigate{1}{\rotatebox[origin=c]{270}{\scriptsize $XX(+)$}} & \ghost{\rotatebox[origin=c]{270}{\scriptsize $XX(-)$}} & \gate{\scriptsize X(+)} & \qw\\
        \lstick{\ket{0}} & \ghost{\rotatebox[origin=c]{270}{\scriptsize $XX(+)$}} & \ghost{\rotatebox[origin=c]{270}{\scriptsize $XX(-)$}} & \ghost{\rotatebox[origin=c]{270}{\scriptsize $XX(+)$}} & \ghost{\rotatebox[origin=c]{270}{\scriptsize $XX(-)$}} &
        \ghost{\rotatebox[origin=c]{270}{\scriptsize $XX(+)$}} &
        \ghost{\rotatebox[origin=c]{270}{\scriptsize $XX(-)$}} & \qw
        & \meter
    }
    }
    \caption{The weight-$6$ stabilizer syndrome extraction circuit (above) and its stabilizer-sliced and fully cancelled ion trap gate compiled circuit (below). Here, $X(\pm)$ corresponds to a $\pm\pi/2$ rotation about $X$, and $XX(\pm)$ is a two-qubit $\chi = \pi/4$ M\o{}lmer-S\o{}renson gate. Note that the slicing in this instance amounts to choosing the directions of the Pauli rotations and M\o{}lmer-S\o{}renson gates.}
    \label{compilation}
\end{figure}
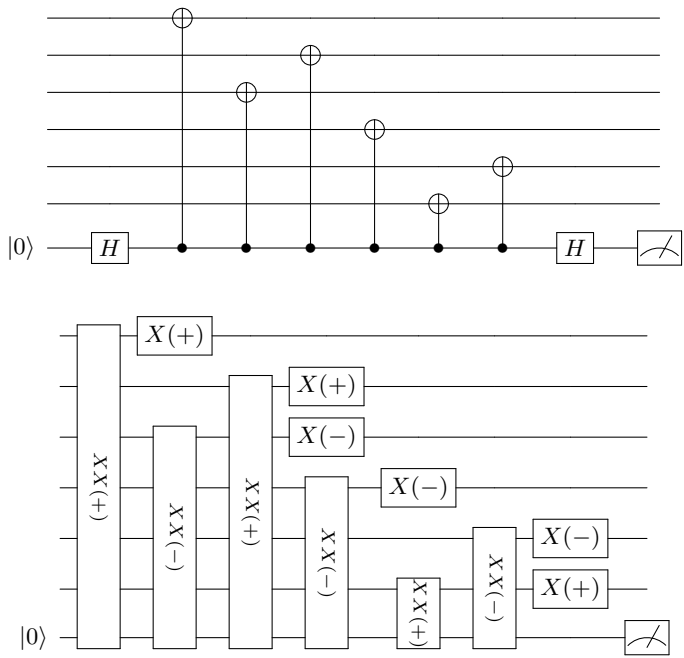

Our next simulation for Bacon-Shor assumes current technologies and implements a more physically grounded error model and gate set. We use the decompositions in Figure \ref{gates} to convert our $6$-body Bacon-Shor stabilizers into ion trap gates, as in Figure \ref{compilation}.
In this model, we consider overrotations on both one- and two-qubit gates, with $1 + \epsilon_2 = (1 + \epsilon_1)^2$ reflecting the quadratic dependence of the two-qubit Rabi frequency on the one-qubit Rabi frequency \cite{MolmerPRL1999}. 

\begin{figure}[ht!]
    \centering
    \includegraphics[width=.9\linewidth]{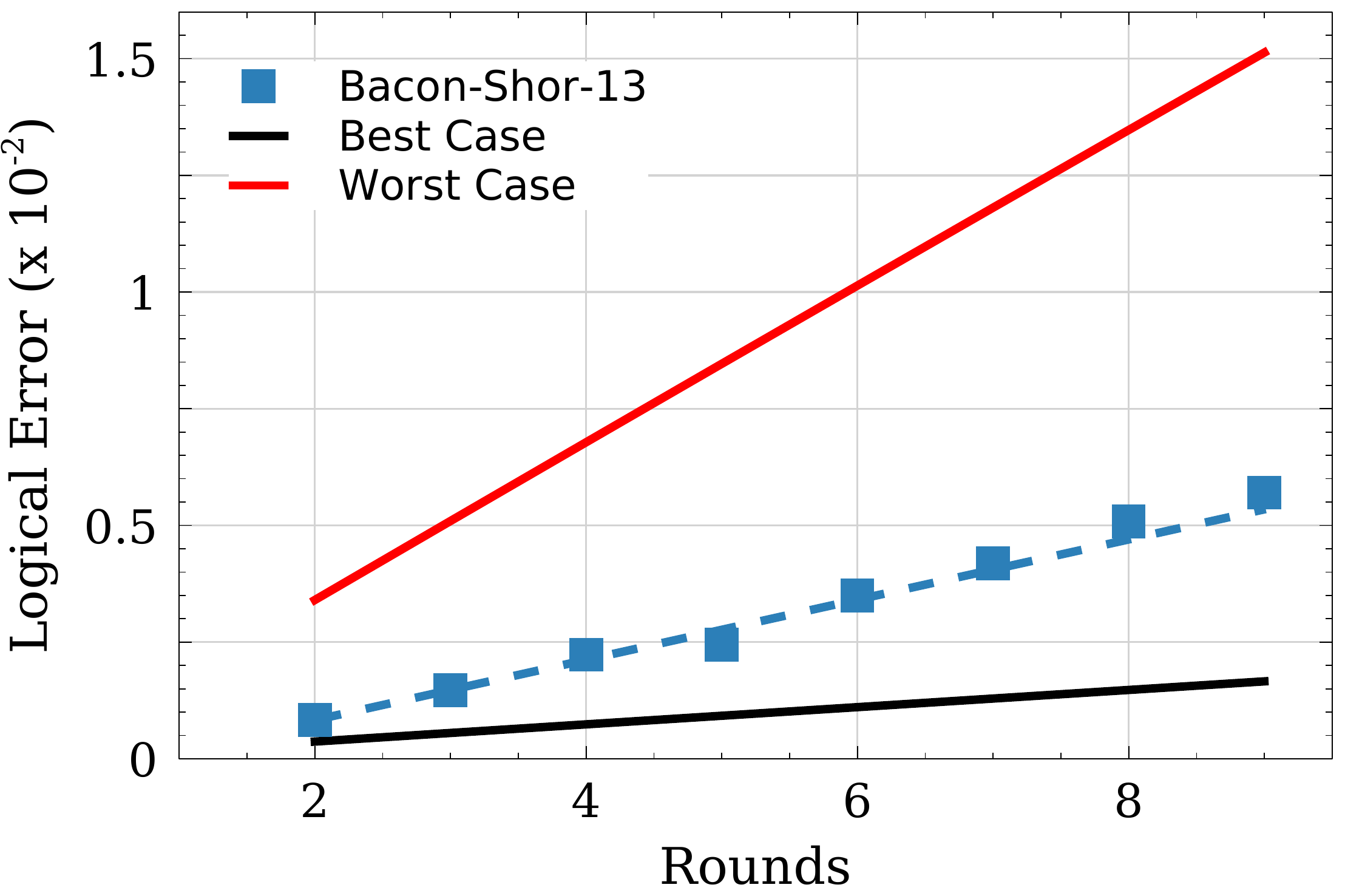}
    \caption{One-sided logical error rates from multiple rounds of error correction using the fully optimized Bacon Shor circuit in the purely coherent case, along with lines indicating error rates for the perfect $+1$ gauge, the gauge featuring the least cancellations, and a linear fit. The effects of gauge decay can be mitigated if one is willing to change their circuit in real time as errors are detected and corrected. Note that the black line is equivalent to Shor's subspace code. Due to the number of postselections required to simulate the full channel, this plot was generated by sampling according to the same error model as Figure~\ref{shor}. }
    \label{multRds}
\end{figure}

In order to make this a more feasible system, we restrict our qubit number and consequently can only measure large stabilizers in parallel. As a result, instead of slicing stabilizers, we are limited to slicing gauges instead. This is an issue as our cancellations rely on the eigenvalue of the operator we are slicing to be $+1$. As a result we can prepare into a $|0\rangle_L$ state with all $X$-gauges being equal to $+1$, but over time, $Z$-errors and corrections will flip these $X$-gauges and lead to a degradation of our cancellations. To understand the impact that this gauge decay has on our system, we consider multiple rounds of error correction in Figure \ref{multRds}.
By occupying an $X$-type gauge we occupy a superposition over all $Z$-type gauge eigenstates. Consequently, there will be no suppression of $Z$-type coherent errors. As a result the error shown in the Figures \ref{shor} is the one-sided infidelity of $|0\rangle_L$. 
\section*{Conclusions}

In both simulations, stabilizer slicing shows marked improvements in the logical error rate, and could be extended to several other codes without modification. We would like to point out a few cases where stabilizer slicing applies in a wider context. 

First, note that our decomposition into two components needn't be symmetric.  Although we have described stabilizer splitting as occurring between two equally-weighted (and in Figure~\ref{stabilizer_slicing}, disjoint) sets of Pauli operators, the only requirement is that $S = S_LS_R$.  

However, to remain experimentally motivated, we wouldn't expect that systematic overrotations between different many-body interactions would be of the same approximate magnitude. While this is physically realistic for the same process mediating the same multi-qubit interaction among different subsets of qubits, processes mediating different weight interactions will likely have different relative overrotations leading to imperfect cancellations.

It is worth noting that our technique can also be thought of in terms of dynamical decoupling~\cite{viola1999dynamical}. Instead of alternating the directions of each section of the stabilizer, you could surround one of the Pauli operators with errors of the opposite type and then have both Pauli operators rotate in the same direction. In this picture the coherent overrotations take the place of the constant background decoherence dynamical decoupling usually works with. These two ideas are equivalent when assuming perfect single qubit gates, however when that approximation is removed the single qubit overrotations cause the higher order terms to not cancel perfectly, in contrast with our method. This decoupling method also results in an inverted stabilizer measurement value, but that can be corrected classically.

Stabilizer slicing also extends to Shor-style syndrome extraction using a Bell state $\ket{\Phi^+}$.  This follows from the Bell state itself satisfying $ZZ\ket{\Phi+} = XX\ket{\Phi^+} = \ket{\Phi^+}$.  By using native multi-qubit interactions, we are already implicitly saving on the circuit-depth of syndrome extraction.  Using such a scheme would allow syndrome-extraction in a single gate-layer.  

The downside of this approach is that, even using Bell states, we may introduce multi-qubit correlated errors due to failures in native multi-qubit gates.  However, for LDPC codes on which these few-body stabilizer interactions may be reasonable, these correlated errors cannot propagate too badly.  In certain cases, such as the hook errors of the surface code, such errors will not lower the effective distance of the code.  Furthermore, for architectures where stochastic depolarizing errors due to gate failure are rare, coherent error mitigation may be well worth the trade-off. 

Lastly, we note that the concept of stabilizer slicing can be applied to any circuit suffering from systematic coherent errors, with varying efficacy.  In particular, preparation circuits with a fixed input may be a good candidate to extend stabilizer slicing beyond syndrome extraction.

In summary, stabilizer slicing is a new and simple technique for suppressing coherent errors in syndrome extraction.  It requires certain experimental capacities but no additional overhead, and dramatically improves the logical fidelity of syndrome extraction with the same-quality physical gates.  Because it requires no additional resources, we hope that even its $2$-body iteration could yield significant benefit in realistic near-term fault-tolerance experiments where systematic coherent error is a dominant factor.  

\section{Acknowledgements}

KRB thanks Thomas O'Brien for useful discussions about quantumsim. The authors also thank Pavithran Iyer, Pak Hong Leung, Leonardo Andreta de Castro and Yukai Wu for helpful conversations.  This work was supported by the Office of the Director of National Intelligence - Intelligence Advanced Research Projects Activity through ARO contract W911NF-16-1-0082, the ARO MURI on Modular
Quantum Systems W911NF-16-1-0349, National Science Foundation Expeditions in Computing award 1730104, and National Science Foundation Phy-1818914.

\bibliographystyle{apsrev}
\bibliography{References}

\section{Supplemental Material}
\setcounter{figure}{6}
Here we have three additional figures that present stabilizer slicing in more realistic contexts. First, we show a $2$-body ion trap Surface-$17$ simulation, where the only benefits are along the boundary of the code in Figure \ref{surface_2_body}. While we do see slight improvements, the stochastic side is artificially bolstered because the majority of single qubit gates are opposed in the unoptimized circuit. The error model for this plot is the same as in Figure 2, only having overrotations on one- and two-qubit gates.
\begin{figure}[ht]
    \centering
    \includegraphics[width = .9\linewidth]{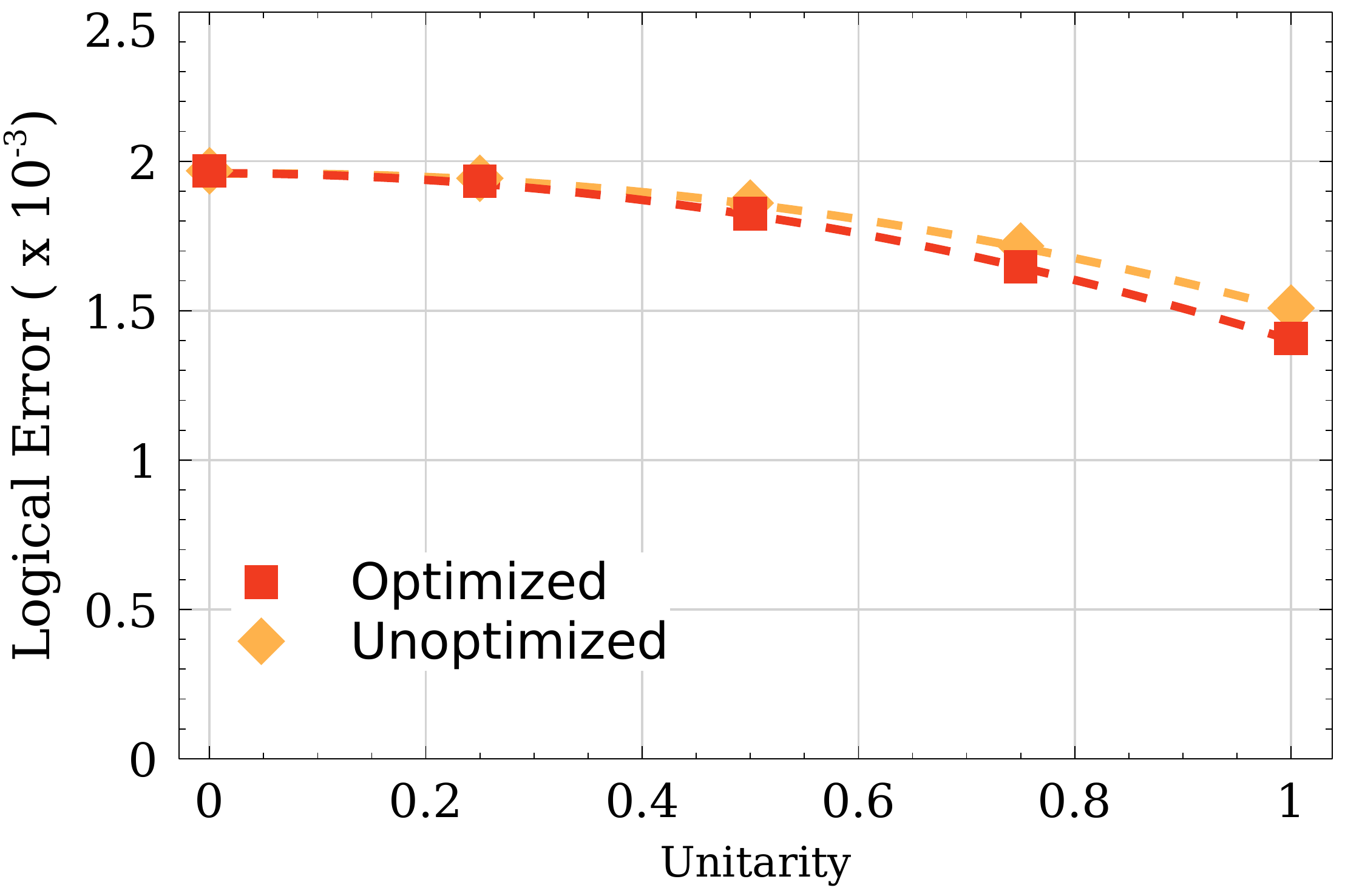}
    \caption{Logical error rates and quadratic fits for Surface-$17$ assuming access only $2$-body ion trap gates both with and without stabilizer slicing. Since we do not have access to $3$-body gates, there are no perfect cancellations for the weight $4$ stabilizers. We still see an improvement, but as the code is scaled up in size, the bulk will dominate the boundary effects and this suppression will be minimal. This plot is generated at a $2$-qubit gate infidelity of $5.0 \times 10^{-4}$, matching Figure 2.}
    \label{surface_2_body}
\end{figure}
\begin{figure}[ht!]
    \centering
    \includegraphics[width = 0.9\linewidth]{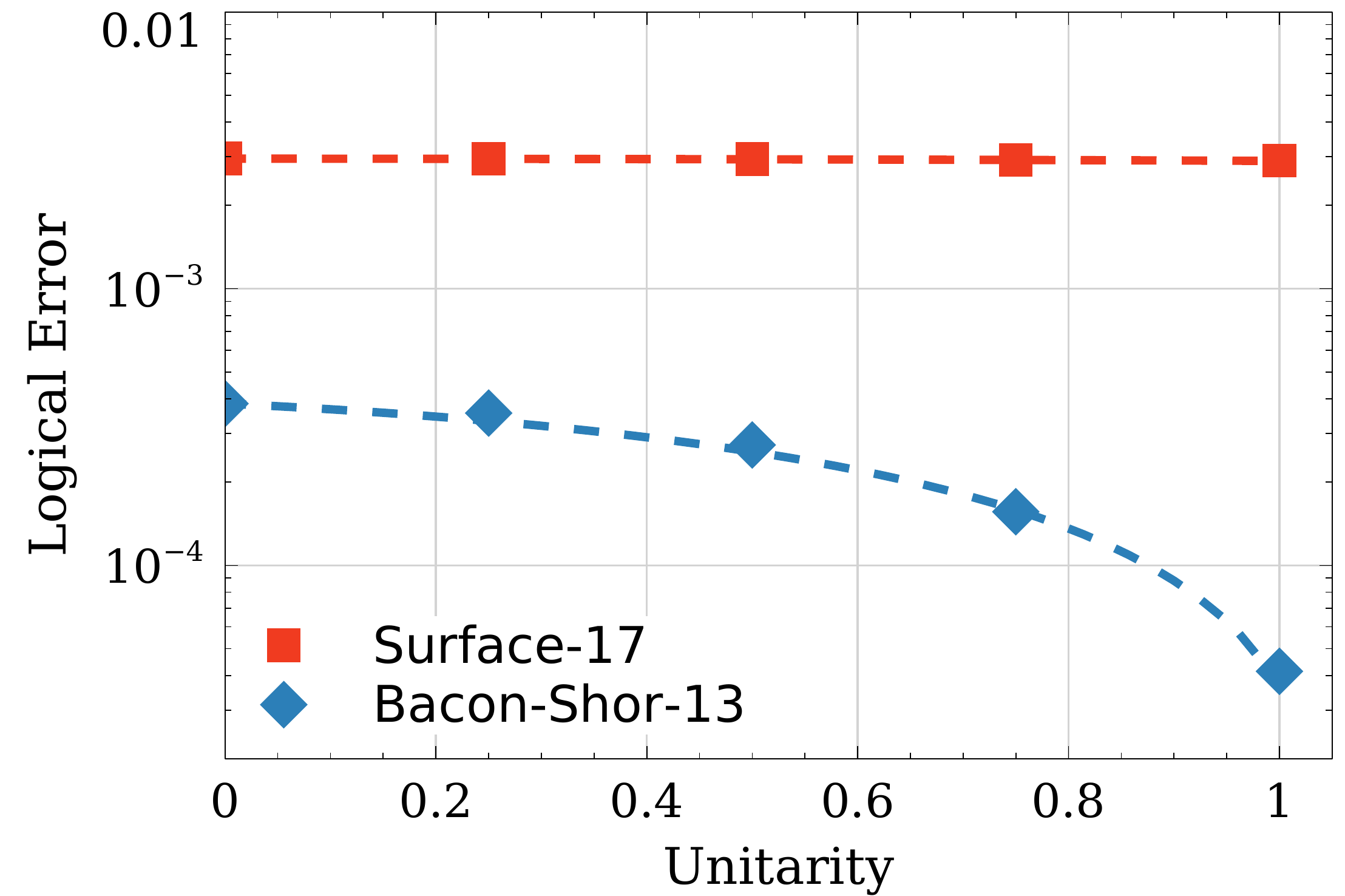}
    \caption{Logical error rates for Bacon-Shor-$13$ and Surface-$17$ after circuit optimization to minimize two-qubit error, and full removal of cancelling single qubit gates. This plot is generated at a $2$-qubit gate infidelity of $1.0 \times 10^{-3}$.}
    \label{fullOpFig}
\end{figure}
In Figure \ref{fullOpFig} we can see the effects of our method in the most realistic context, with fully optimized and cancelled circuits. In these circuits, stabilizer slicing is implemented to the extent that our $2$-body operations allow, and all single qubit gates which can be cancelled are removed entirely from the system. These circuits are implemented under a physically motivated noise model which includes not only overrotations, in both coherent and Pauli twirled forms, but also single qubit dephasing on each qubit involved in a gate, with a magnitude dependent on the one- and two-qubit gate times. Once dephasing is brought into the picture, Surface-$17$ stops performing better with increasing coherence, but the Bacon-Shor-$13$ code has more cancellations and so still improves with coherence.
\begin{figure}[b]
    \centering
    \includegraphics[width=0.9\linewidth]{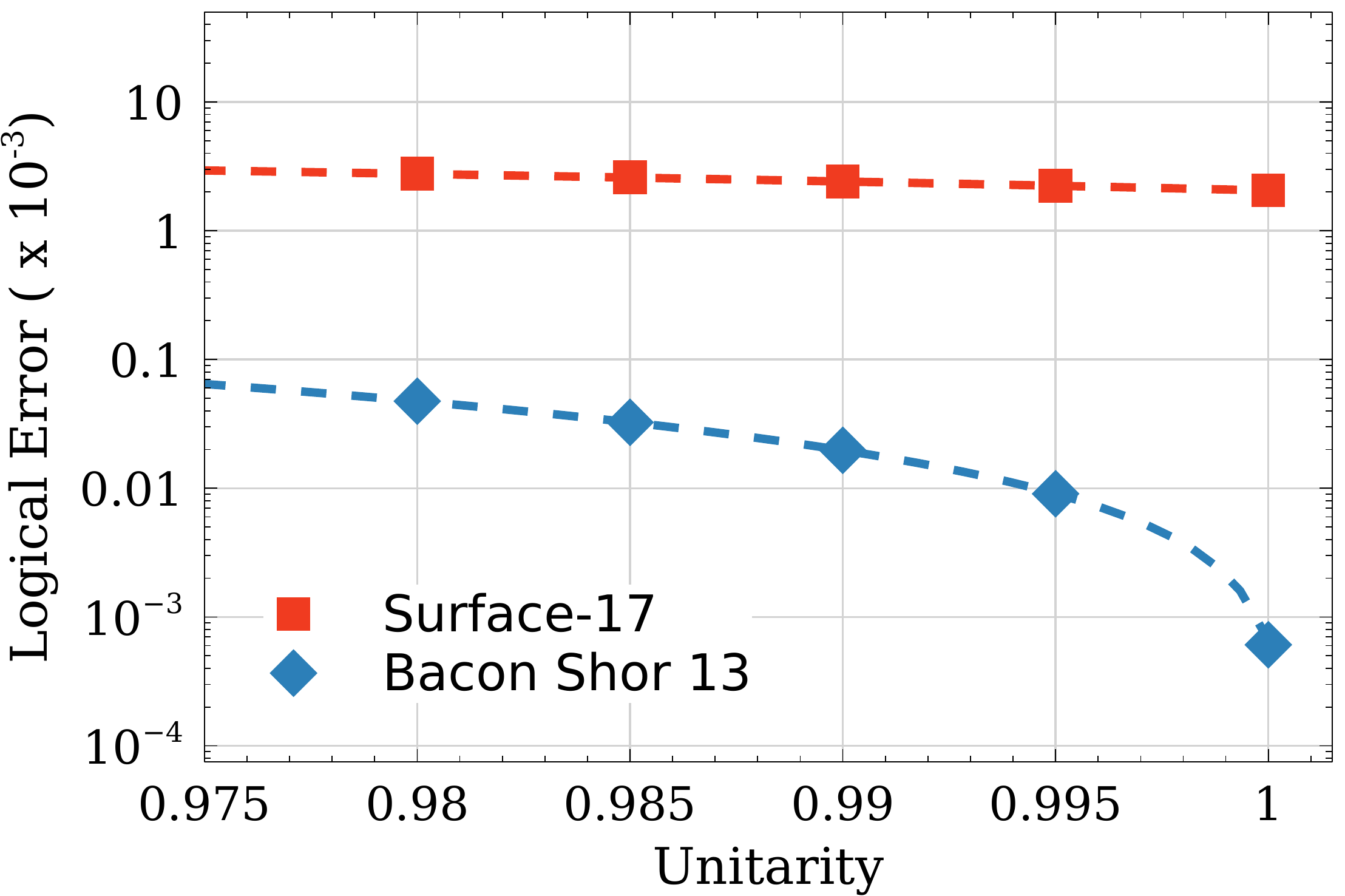}
    \caption{Logical error rates for fully optimized and cancelled codes with SK1 implemented to reduce single qubit overrotation error. The error model only has one- and two-qubit overrotations. This plot is generated at a $2$-qubit gate infidelity of $1.0 \times 10^{-3}$.}
    \label{sk1fig}
\end{figure}
Lastly, in Figure \ref{sk1fig} we apply the same circuits but with the SK1 protocol implemented on all single qubit gates. We apply the same overrotation error model to all the gates in the circuit, including the large arbitrary axis rotations that are used in SK1, and as a result the stochastic errors are significant. We only plot these simulations in the regime where SK1 improved the single qubit gate fidelities, which is above $\kappa = 0.9847$ for our error model.
\clearpage

\end{document}